\documentclass[aps, pra, twocolumn]{revtex4}
\usepackage{graphicx}
\usepackage{tikz}
\usepackage[caption=false]{subfig}

\begin{document}

\title{Disorder in the spectral function of a qubit ensemble}
\author{Samuel L Smith}
\email{sls56@cam.ac.uk}

\author{Alex W Chin}
\affiliation{Theory of Condensed Matter, Cavendish Laboratory, University of Cambridge}

\begin{abstract}
The impact of disorder in the vibrational bath of an ensemble of open quantum systems is explored, arising either from variation in the overall coupling strength or from uncertainty in the shape of the environment spectral function. Such disorder leads to an additional source of decoherence of subsystem ensembles, due to variation in the reorganization energy. Additionally, vibrational disorder induces a shift in the oscillation frequency of the coherence between ground and excited states. This shift is temperature dependent, and in the high temperature limit $\delta \omega \propto T$. This latter finding could be of particular relevance to biological systems exhibiting long lived coherences in highly disordered environments, where temperature-dependent quantum beats have been observed.
\end{abstract}

\maketitle

\section{Introduction}
The decay of quantum coherences is an active topic of research across a wide range of fields in physics, chemistry and increasingly biochemistry \cite{leggett1987dynamics,petruccione2002theory,weiss1999quantum}. An important source of decoherence arises from interactions between the subsystem of interest and a vibrational heat bath \cite{leggett1987dynamics,petruccione2002theory,weiss1999quantum,mohseni2014quantum}. The vibrational bath can be characterized by a continuous spectral density, $J(w)$, which quantifies the density of vibrational states with frequency $w$ in the range $0<w<\infty$, weighted by the strength of their coupling to the subsystem. While dissipative processes are often controlled by the finite-frequency structure of the spectral function, coherence losses due to \emph{pure dephasing} are primarily controlled by the behaviour of this spectral function as it approaches zero frequency \cite{weiss1999quantum, petruccione2002theory,giraldi2013survival}. A high density of low frequency modes is expected to drive a rapid decay in coherences, while an absence of such modes may enable partial coherence to survive over long timescales.

The pure dephasing properties of general, power-law spectral functions have been thoroughly investigated and have found important applications in quantum information and metrology \cite{wilhelm2008quantum, chin2012quantum}. Simple pure dephasing models, such as the independent boson model, can be solved exactly \cite{mahan2000many}. Recently, it was proposed that long-lasting quantum coherences observed in biological photosynthetic systems could be stabilized by suppressed low frequency spectral densities, enabling long lived quantum coherences between excitonic states to survive at physiological temperatures. Kreisbeck and Kramer \cite{kreisbeck2012long} have investigated this with advanced numerical methods for the Fenna-Matthews-Olson (FMO) complex, finding that the super-Ohmic behaviour of experimentally derived spectral functions for FMO greatly extend exciton-exciton coherences without compromising efficient energy transfer \cite{kreisbeck2012long}. A number of other theories, invoking highly structured spectral densities typical of protein-bound chromophores, have also been presented \cite{chin2013role,christensson2012origin,ishizaki2009theoretical,kolli2012fundamental,lambert2013quantum,mohseni2014quantum,tiwari2013electronic}.

Biological systems are known to exhibit a high degree of static, structural and dynamic disorder \cite{olbrich2010quest,shim2012atomistic}, yet typically their vibrational environments are modeled by well-defined spectral functions. The goal of this paper is to explore, within a simple model, the effect of incorporating a random component to the mean spectral function. Though this problem is inspired by quantum biology, it has potential relevance across a much wider range of applications.
\begin{figure}[b]

\begin{tikzpicture}[scale=0.9]

	\draw[->] (-0.2,0) -- (6,0) node[right] {$\textbf{R}$};
	\draw[->] (0,-0.2) -- (0,4) node[left] {$E$};

	\draw[color=red, domain = 0.5:4.3] plot (\x, {0.3*(\x-2.5)^2 + 0.5});
	\draw[color=blue, domain = 2.064:5.936] plot (\x,{0.4*(\x-4)^2 + 2.5});
	
	\draw[->] (2,0.5)  -- (2,3.4);
	\draw[<-](2.5,2.5) -- (2.5,3.4);
	\draw[->] (4.5,0.5) -- (4.5,2.5);

	\draw (2,0.5) -- (2,2.5) node[left] {$E_1$};
	\draw (2.5,2.5) -- (2.5,3.05) node[left] {$\Delta$};
	\draw (4.5,0.5) -- (4.5,1.5) node[right] {$E_1 - \Delta$};

	\draw[dotted] (1,3.4) -- (5,3.4);
	\draw[dotted] (3,2.5) -- (5.5,2.5);
	\draw[dotted] (1.5,0.5) -- (5.5,0.5);

	\draw[dashed, color=blue, domain = 2.9:5.1] plot(\x,{0.4*(\x-3.5)^2 + 3});
	\draw[dashed, color=blue, domain = 2.135:6] plot(\x,{0.4*(\x-4.371)^2 + 2});
	\draw[dotted] (3,2) -- (5.5,2);
	\draw[dotted] (3,3) -- (5.5,3);
	\draw[<->] (5.3,3)--(5.3,2) node[right] {$2\sigma_{\Delta}$};

\end{tikzpicture}

\caption{During direct excitation, the two energy levels are separated by an energy gap $E_1$. The vibrational states then respond; lowering the energy of the excited state to $E_1 - \Delta$, where $\Delta$ labels the reorganization energy. In the presence of vibrational disorder, a distribution of reorganization energies emerges, shown here with standard deviation $\sigma_{\Delta}$. Realizations of the system with smaller reorganization energies experience less pure dephasing, leading to an apparent shift in oscillation frequency.}
\label{fig:3}
\end{figure}
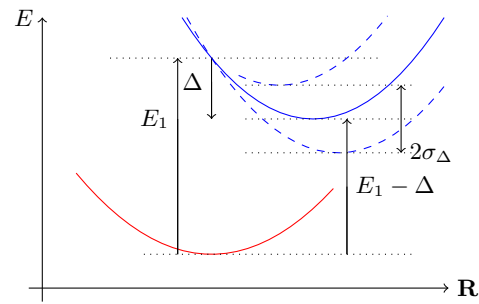

\section{A simple model, the two level system}
Consider the independent boson model \cite{mahan2000many,weiss1999quantum}, described by the Hamiltonian $H = H_S + H_B + H_{SB}$, and represented schematically in figure \ref{fig:3}.
\begin{eqnarray}
H_S & = & |0><0| + E_1|1><1|, \\
H_B & = & \sum_k \omega_k a_k^\dagger a_k, \\
H_{SB} & = &     \left[\sum_k (g_k a_k^\dagger + g_k^*a_k) \right] |1><1|. 
\end{eqnarray}
$H_S$ represents a simple two level system. The upper level is coupled to a set of vibrations, represented by bosonic oscillators. We can characterize the vibrational states by the continuous spectral density function $J(w) = \sum_k |g_k|^2 \delta(w-w_k)$ \cite{weiss1999quantum}.

Imagine coherently exciting the system to the upper state at time $t=0$. Since ground and excited states are decoupled, the population in the upper state cannot decay; this is represented by the diagonal component of the density matrix $\rho_{11}(t) = 1$. At time zero the coherence between the ground and excited states will also be unity, represented by the off diagonal component $\rho_{01}(0) = 1$. We are interested in the evolution of this coherence over time.

\begin{figure}[h]
\centering
\includegraphics[scale = 0.6]{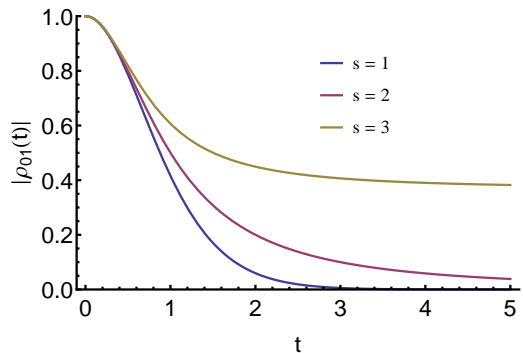}
\caption{Coherence decay for spectral functions with $s = 1$, $2$ and $3$. The reorganization energy $\Delta$ is fixed to $0.1$, and we have chosen the energy scale $w_c = 1$. We have used the high temperature limit $coth(\beta w/2) \sim 2k_BT/w$, and chosen $k_BT=10$.}
\label{fig:1}
\end{figure}

Since the system was excited abruptly at $t=0$, we are free to assume that the bath and the subsystem are initially unentangled; at $t=0$ the bath lies in the thermal state $e^{-\beta H_B}/Tr[e^{-\beta H_B}]$. Under such assumptions, the evolution of the system is known to obey \cite{petruccione2002theory}
\begin{eqnarray}
\rho_{12}(t)& = &\rho_{12}(0)e^{-i E_1 t}e^{-\gamma(t)},  \label{eq:1} \\
\gamma(t)& = & \int_0^{\infty} dw \; J(w)f(w, t), \\
f(w)& = & \frac{[1-cos(wt)]}{w^2} \, coth(\beta w/2) \nonumber \\  
&& + \, i  \, \frac{[sin(wt)-wt]}{w^2}. \label{eq:2}
\end{eqnarray}
Usually we are only interested in the coherence decay, and so the imaginary terms in the exponentials above are often neglected. As discussed earlier, the coherence decay is dominated by the behaviour of $J(w)$ at low frequencies. To demonstrate this we characterize the low frequency components of the spectral function within the standard form $J(w) = \alpha w_c^{1-s} w^s e^{-w/w_c}$ \cite{leggett1987dynamics,weiss1999quantum}. The behaviour of the coherence decay depends critically on the parameter s. At any finite temperature, if $0 < s < 2$ then the coherence $|\rho_{12}(t)|$ will decay exponentially to zero in the long time limit \cite{weiss1999quantum}. The rate of this decay is linear in the spectral coefficient $\alpha$, and the decay becomes arbitrarily slow as s approaches 2 \cite{weiss1999quantum,giraldi2013survival}. Meanwhile if $s>2$ then the coherence decays initially, before settling at a finite value $|\rho_{12}(\infty)| > 0$ \cite{weiss1999quantum}. The evolution of coherence for $s=1$, $2$ and $3$ is exhibited in figure \ref{fig:1}. Consequently, we might hope to suppress the decoherence of our subsystem by engineering a spectral density which suppresses the growth of the spectral density at low frequency; such that the coefficient $\alpha$ corresponding to any component with $s<2$ is small. This idea has been proposed in a range of contexts, most recently as a means of sustaining coherences within photosynthetic systems \cite{kreisbeck2012long}.

The analysis above has considered the spectral density to be a single well defined function. In practice, most experiments are performed over an ensemble of systems. Additionally in the photosynthetic context, excited states coherences within a single system must be preserved between the different molecular sites contributing to transport. Each site will be coupled to a different vibrational bath. Since any ensemble will exhibit a range of spectral densities; we must quantify the effect spectral disorder has on the coherence $\rho_{01}(t)$.

In the above we neglected the oscillatory terms. For later convenience we note that, in the absence of spectral disorder, the complex term oscillates as $Exp[-i(E_1 - \alpha \Gamma(s) w_c)t]$ in the long time limit. We may identify the reorganization energy $\Delta = \alpha \Gamma(s) w_c$. We also note that $\Gamma(3) = 2$.

\section{Disorder in the spectral coefficient}

\begin{figure}[t]
\subfloat{(A)\includegraphics[scale = 0.7]{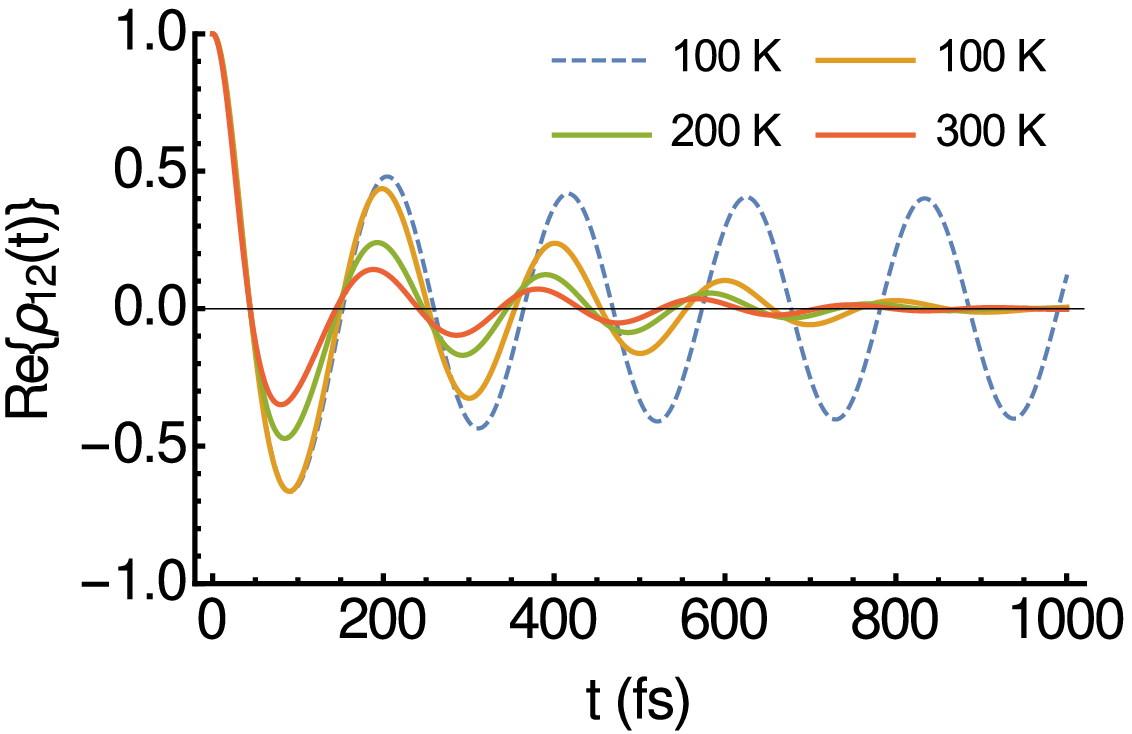}} \\
\subfloat{(B)\includegraphics[scale = 0.34]{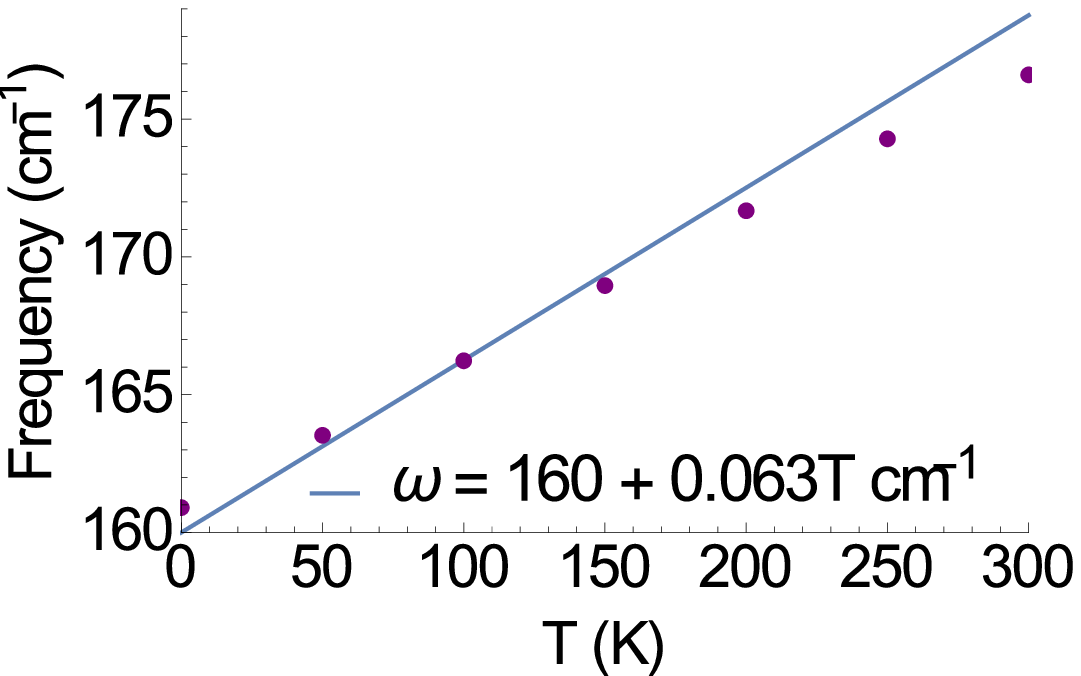}} 
\subfloat{  (C)\includegraphics[scale = 0.34]{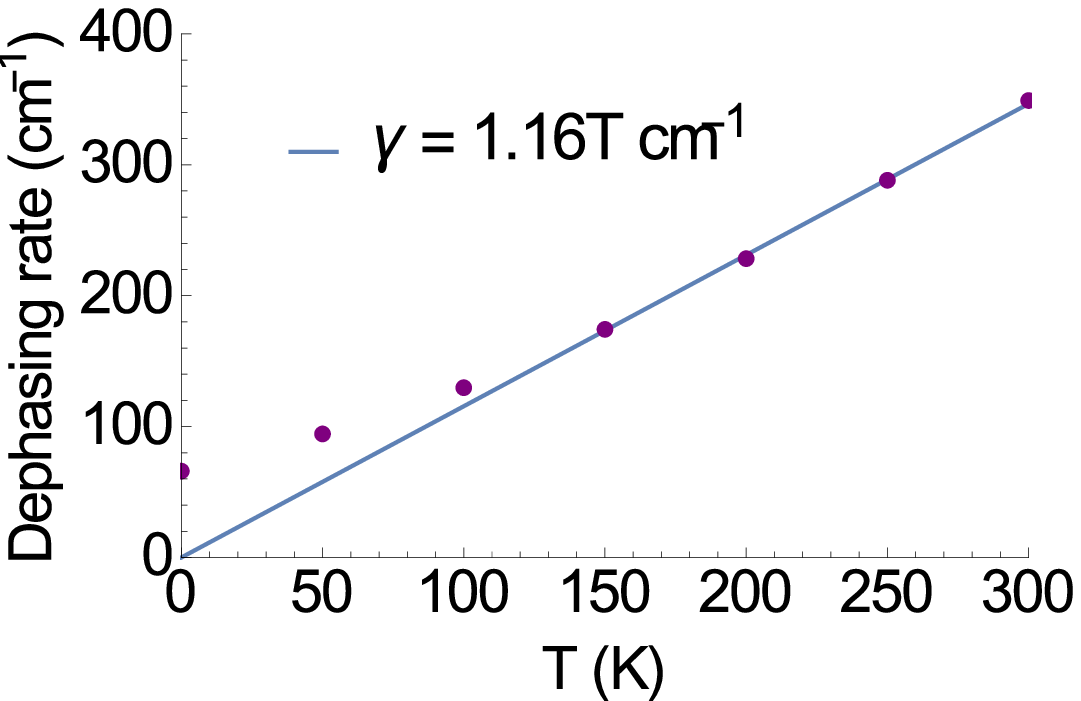}} \\
\caption{(A) Coherence oscillations at a range of temperatures. We fix $w_c = w_p/3 =$ 50 cm$^{-1}$, $\Delta =$ 40 cm$^{-1}$, $\sigma_{\Delta} =$ 15 cm$^{-1}$, and the underlying energy gap $E_1 =$ 200 cm$^{-1}$. For the dashed curve, the disorder $\sigma_{\Delta} =$ 0. (B/C) We fit these curves to the function $cos(\omega t)\, e^{-\gamma t}$, and extract estimates for the frequency shift and dephasing rate. We predict the frequency shift using eqn. 19, while the dephasing rate is described by a line of best fit. We note that for the parameters considered here, the frequency shift is similar to that observed in Ref. \cite{panitchayangkoon2010long}. but the dephasing rate is roughly twice as large.}
\label{fig:2}
\end{figure}

For the remainder of this paper we consider the special case $s=3$, which is a common form of spectral function arising from the coupling to bulk phonons \cite{weiss1999quantum},  and we choose our energy scale such that $w_c=1$. Thus the  mean spectral density $\bar{J}(w) = \alpha w^3 e^{-w}$. Furthermore, we enter the high temperature limit $\beta w_c \ll 1$. In this limit we may expand $coth(\beta w/2) \to 2k_BT/w$. Recalling equations \ref{eq:1}-\ref{eq:2} we define
\begin{eqnarray}
\varphi(t) & = & \frac{1}{\alpha} \int_0^{\infty} dw \bar{J}(w) f(w,t) \\
           & = & \varphi_R(t) + i\varphi_I(t).
\end{eqnarray}
Under our approximations
\begin{eqnarray}
\varphi_R(t) & = & 2k_BT\frac{t^2}{1+t^2}, \\
\varphi_I(t) & = & -2t\left[1-\frac{1}{(1+t^2)^2} \right].
\end{eqnarray}
The simplest way to introduce disorder to our system is to allow the spectral coefficient $\alpha$ to vary, while holding the spectral distribution constant.  This situation, where the reorganization energies of different, spatially separated pigments varies was proposed for pigment-protein complexes in Refs. \cite{olbrich2011theory,shim2012atomistic}. To do this, we average across the probability distribution of $\alpha$, denoted by $P(x)$.
\begin{eqnarray}
\rho_{12}(t)  & = & e^{-E_1 t} \phi(t), \\
\phi(t) & = & \int_0^{\infty} dx P(x) e^{-x \varphi(t)}.
\end{eqnarray}
The coherence evolution will depend on the choice of probability distribution $P(x)$. For simplicity, we assume that $P(x)$ is Gaussian distributed about the mean value $\alpha$ with standard deviation $\sigma$.
\begin{equation}
\phi(t) = f(\alpha,\sigma) \int_0^{\infty} dx  e^{-(x-\alpha)^2/2\sigma^2}e^{-x\varphi(t)}. \label{eq:9}
\end{equation}
The prefactor $f(\alpha, \sigma)$ ensures normalization. Equation \ref{eq:9} has an exact solution in terms of the error function. This solution is used in figure \ref{fig:2}, where we plot the coherence oscillation for fixed $\sigma$ at a range of temperatures. To understand the behaviour physically we assume that $\sigma \ll \alpha$ and extend the lower limit of integration to $-\infty$ to obtain
\begin{eqnarray}
\phi(t) & = & e^{-\alpha \varphi(t)+\sigma^2 \varphi(t)^2 /2} \label{eq:4} \\
        & = & Exp[- \alpha \varphi_R - \sigma^2(\varphi_I^2 - \varphi_R^2)/2]\times  \nonumber \\
            && Exp[-i (\alpha- \sigma^2 \varphi_R) \varphi_I] \label{eq:3}.
\end{eqnarray} 
Extending the lower limit of integration requires $\sigma^2 \varphi_R \ll \alpha$; this ensures that the real part of $f(\omega)$ induces a coherence decay as expected. Thus the result above can only be valid for spectral densities with $s>2$, since $\varphi_R(t)$ must be bounded. Neglecting this small term we obtain the coherence decay
\begin{eqnarray}
 && |\rho_{01}(t)|  =  Exp[-\alpha \varphi(t) - \sigma^2 \varphi_I(t)^2/2] \\
               = \, && Exp\left[-\frac{2\alpha k_BT t^2}{1+t^2} - 2\sigma^2 t^2 \left(1- \frac{1}{(1+t^2)} \right)^2\right].
\end{eqnarray}
In the absence of disorder the coherence stabilizes at a minimum value, $|\rho_{12}(\infty)| = Exp(-2\alpha k_B T)$. Introducing disorder causes the coherence to decay to zero on a timescale $1/\sigma$. In hindsight, the origin of this decay is clear. The disorder in the spectral function causes the reorganization energy to become disordered, at times $t > 1$ this introduces a static energetic disorder into the transition energy, which drives a coherence decay over the ensemble. This dephasing term is suppressed for $t < 1$, since the vibrational bath takes a finite time to relax. In the case considered, the reorganization energy is also Gaussian distributed with standard deviation $\sigma_{\Delta} = 2\sigma$.

More striking however, is the oscillating term in Eq. \ref{eq:3}. Denoting $\omega_B t = (\alpha - \sigma^2 \varphi_R)\varphi_I$, and entering the long time limit $t \gg 1$ we find
\begin{eqnarray}
\omega_B & = & -2(\alpha - 2\sigma^2k_BT) \\
	 & = & -\Delta + 4\sigma^2k_BT 
\end{eqnarray}
The disorder has introduced a small shift in the oscillation frequency of the coherence, which grows linearly as temperature increases. Under our earlier approximations, this expression fails when $2 \sigma^2 k_BT  \sim \alpha$. 

This shift has an intuitive origin. When we introduce disorder to our spectral function, the variation in the real coherence decay (driven by $\varphi_R)$ is correlated to variation in the reorganization energy $\Delta$. Consequently the average over oscillation frequencies becomes a weighted average favouring higher frequencies (smaller reorganization energies), and the apparent oscillation frequency measured experimentally differs from the true mean value. The shift grows with temperature, since $\varphi_R \propto k_BT$. Reinstating $w_c$ we find that $d \omega_B/d T = \sigma_{\Delta}^2 k_B/w_c^2$, or $0.7 \, (\sigma_{\Delta}/ w_c)^2 \, \text{cm}^{-1}/\text{K}$. $\sigma_{\Delta}$ labels the standard deviation of the reorganization energy. We note that $\omega_B$ is independent of time when $t \gg 1$ because $\varphi_R(t)$ is bounded, this would not be the case if $s\leq 2$. Although this shift is small, it could potentially be observed experimentally in appropriate systems. Though we have considered a specific spectral function here, a similar effect may arise whenever a system exhibits significant disorder in the density and coupling strength of low frequency modes. 

Indeed such an effect could underlie the increase of beating frequency with temperature observed by Panitchayangkoon \textit{et al} \cite{panitchayangkoon2010long}. Panitchayangkoon tentatively observed a shift of $\sim$5 cm$^{-1}$ between 77 and 150K, in the coherence between excitons 1 and 3 of the FMO complex. Accounting for such a shift within our model requires that $\sigma_{\Delta}\sim 0.3 w_c = 0.1 w_p$, where $w_p$ labels the frequency of the peak in the spectral function. There remains a great deal of uncertainty regarding the spectral function of the FMO complex \cite{olbrich2011theory}, and to our knowledge there is no published work on the statistics of spectral function disorder within an ensemble. Most studies have not used the simple cubic spectral function considered here. However, as a rough estimate we note that the smooth, low frequency part of these spectral functions typically have values for $w_p$ in the range 150-250 cm$^{-1}$. This implies a standard deviation $\sigma_{\Delta} \sim$ 15-25 cm$^{-1}$. Since typical values for $\Delta$ lie in the range 20-60 cm$^{-1}$ \cite{adolphs2006proteins,kreisbeck2012long,olbrich2011theory,shim2012atomistic}, this suggests that the ratio $\sigma_{\Delta}/\Delta \sim$ 0.25-1.25. Given the limited data available, we conclude that spectral function disorder could plausibly generate a shift on the same order of magnitude as that observed experimentally, though many other factors would need to be assessed in order to make any definite assignment. 

\section{Disorder in the Spectral Distribution}
The previous section illustrated the key effects of disorder on the coherence evolution within a simple model. Here we demonstrate that the same behaviour is observed when a more realistic continuously varying spectral disorder is considered.

We consider the spectral function $J(\omega) = \bar{J}(\omega) + \sigma_J(\omega)$. $\sigma_J(\omega)$ represents some random disorder at the frequency $\omega$, and is correlated over a frequency range $\xi \ll 1$. Averaging over disorder realizations implies
\begin{eqnarray}
\rho_{12}(t) & = & e^{-i(E_1t+ \alpha \varphi_I(t))}e^{-\alpha \varphi_R(t)} \phi(t), \label{eq:11} \\
\phi(t) & = & \langle Exp[\int_0^{\infty} dw \sigma_J(w) f(w)] \rangle.
\end{eqnarray}
As previously we take $\bar{J}(w) = \alpha w^3 e^{-w}$, and for simplicity we take $\sigma_J(w) = \bar{J}(w)\sigma(w)/\alpha$. We will assume that the probability distribution for $\sigma$ is symmetric about $0$, which implies that $\sigma(w)$ is a random number bounded between $\pm \alpha$. We may express $\phi(t)$ as
\begin{equation}
\phi(t) = \int_{-\infty}^{\infty} \int_{-\infty}^{\infty} dx dy P(x,y) e^{2k_BTx+iy}.
\label{eq:5}
\end{equation}
$P(x,y)$ represents the probability distribution of the two correlated variables
\begin{eqnarray}
x & = &  \int_0^{\infty} dw e^{-w} (1-cos(wt)) \sigma(w), \\
y & = &  \int_0^{\infty} dw e^{-w} w(sin(wt)-wt) \sigma(w).
\end{eqnarray}
At this point we note that both x and y have zero mean. Furthermore both are effectively described by a sum over $O(1/\xi)$ uncorrelated random components. Consequently so long as $\xi$ is small we may apply the central limit theorem and describe $P(x,y)$ via the normal distribution of two correlated variables,
\begin{eqnarray}
&& P(x,y) = \frac{1}{2\pi \sigma_x \sigma_y \sqrt{1 - \rho^2}} \times \nonumber \\
&& Exp[\frac{1}{2(1-\rho^2)} \left( \frac{x^2}{\sigma_x^2} + \frac{y^2}{\sigma_y^2} - \frac{2\rho xy}{\sigma_x \sigma_y} \right) ].
\end{eqnarray}
In the expression above, $\sigma_x^2 = <x^2>$, $\sigma_y^2 = <y^2>$, and $\rho = \sigma_{xy}/(\sigma_x \sigma_y)$ where $\sigma_{xy} = <xy>$. Inserting $P(x,y)$ into equation \ref{eq:5} we find
\begin{equation}
\phi(t) = e^{-(\sigma_y^2 - 4k_B^2T^2 \sigma_x^2)/2} e^{i4k_BT \sigma_{xy}}. \label{eq:8}
\end{equation}
Note that in equation \ref{eq:5} we extended the limits of the integrals over x and y to $\pm \infty$, this is only valid if $ \sigma_x^2 \ll \alpha/k_BT$. In the appendix we show that
\begin{eqnarray}
\sigma_x^2(t)  =  \beta^2 \xi && \int_0^{\infty} dw e^{-2w} (1-cos(wt))^2, \\
\sigma_y^2(t)  =  \beta^2 \xi && \int_0^{\infty} dw e^{-2w} w^2(sin(wt)-wt)^2, \\
\sigma_{xy}(t) =  \beta^2 \xi && \int_0^{\infty} dw e^{-2w} w \times \nonumber \\
                                              &&(1-cos(wt))(sin(wt) - wt).
\end{eqnarray}
The random variable $\sigma(w)$ has been extracted and replaced by its variance $\beta^2$.  Additionally, all three variances are proportional to the correlation length $\xi$. Since $\sigma$ is bounded by $\pm \alpha$, $\beta < \alpha$. The exact solution of these three integrals is given in the appendix, here we simply note that in the long time limit $t \gg 1$, $\sigma_x^2 = 3\beta^2 \xi/4$, $\sigma_y^2(t) = 3t^2\beta^2 \xi/4$, and $\sigma_{xy}(t) = -t\beta^2 \xi/4$. Thus at long times
\begin{equation}
\phi(t) \approx Exp[-\frac{3 \beta^2 \xi t^2}{8}] Exp[-i \beta^2 \xi k_BT t]. \label{eq:10}
\end{equation}
We insert equation \ref{eq:10} into equation \ref{eq:11} to describe the coherence oscillations at long times. Just like in the previous section, the spectral disorder drives the coherence amplitude to zero in the long time limit, and a temperature dependent shift in the apparent oscillation frequency of the coherence is observed. As before, we may obtain the standard deviation of the reorganization energy, $\sigma_{\Delta} = \beta \sqrt{3 \xi/4}$, and the shift in the coherence frequency, $\omega_B = -\Delta + \beta^2 \xi k_B T$. Reinserting $w_c$, we find $d \omega_B /dT = 4 \sigma_{\Delta}^2 k_B/3 w_c^2$.

\section{Conclusions}
In this work we investigated the effect of disorder in the vibrational spectral function of an ensemble of two level systems. We find that disorder introduces an additional coherence decay, since it introduces disorder in the reorganization energy of the excited state. This additional source of decoherence is suppressed at early times, as the vibrational bath exhibits a finite relaxation timescale. As well as this coherence decay, the apparent oscillation frequency of the coherence is slightly increased. This shift occurs because the decoherence of a single realization of the system is correlated to its reorganization energy. The shift increases with temperature, an effect which may be observed experimentally in ultrafast spectroscopies \cite{panitchayangkoon2010long}.

\begin{acknowledgements}
We acknowledge funding from the Winton Programme for the Physics of Sustainability, and would also like to thank John Biggins for useful discussions.

\end{acknowledgements}

\appendix*

\section*{Appendix}
We wish to find the variances and covariance of
\begin{eqnarray}
x & = &  \int_0^{\infty} dw e^{-w} (1-cos(wt)) \sigma(w), \nonumber \\
y & = &  \int_0^{\infty} dw e^{-w} w(sin(wt)-wt) \sigma(w). \nonumber
\end{eqnarray}
$\sigma(w)$ represents a random function correlated over an interval $\xi$. Since the correlation length $\xi \ll 1$ is short, we may approximate
\begin{eqnarray}
x & \approx & \xi \sum_n e^{-n\xi} (1-cos(n\xi t)) \sigma_n, \nonumber \\
y & \approx & \xi \sum_n e^{-n\xi} n\xi (sin(n\xi t) -n\xi t) \sigma_n. \nonumber
\end{eqnarray}
Note that we have explicitly imposed correlation over a frequency scale $\xi$; $\sigma_n$ now refers to independent uncorrelated random numbers. All these random numbers share a common variance $<\sigma_n^2> = \beta^2$. As an example we consider the covariance $<xy>$, this is given by
\begin{eqnarray}
<xy> & \approx & \beta^2 \xi^2 \sum_n e^{-2n\xi} n\xi (1-cos(n\xi t)) (sin(n\xi t) - n\xi t) \nonumber \\
     & = & \beta^2 \xi \int_0^{\infty} dw e^{-2w} w (1-cos(wt))(sin(wt)-wt). \nonumber
\end{eqnarray}
Following this procedure
\begin{eqnarray}
<x^2> & = & \beta^2 \xi \int_0^{\infty} dw e^{-2w} (1-cos(wt))^2, \nonumber \\
<y^2> & = & \beta^2 \xi \int_0^{\infty} dw e^{-2w} w^2(sin(wt)-wt)^2. \nonumber
\end{eqnarray}
Performing the three integrals, we obtain
\begin{eqnarray}
<x^2> & = & \beta^2 \xi \frac{3t^4}{4(4+5t^2 + t^4)}, \nonumber \\
<y^2> & = &  \frac{\beta^2 \xi t^2}{8} \left( 6 + \frac{768(t^2-4)}{(t^2 + 4)^4} + \frac{6+3t^2+t^4}{(t^2+1)^3} \right), \nonumber \\
<xy> & = & -\beta^2 \xi \frac{t^5(120 + 106t^2 + 14t^4+t^6)}{4(t^2+1)^2(t^2+4)^3}. \nonumber
\end{eqnarray}

\bibliography{library}

\end{document}